\documentstyle[11pt,aasms4]{article}

\tighten
\received{May 28th, 1998}
\accepted{September 21th, 1998}
\slugcomment{To appear in Astrophysical Journal}

\def\etal{{\it et al.}\ }

\def\msun{{\rm\,M_\odot}}

\def\s-1{{\rm\,s^{-1}}}

\def\spose#1{\hbox to 0pt{#1\hss}}
\def\C3H2{{\rm\,\rm C_3H_2}}
\def\NH3{{\rm\,\rm NH_3}}
\def\HOCO+{{\rm\,\rm HOCO^+}}
\def\lta{\mathrel{\spose{\lower 3pt\hbox{$\mathchar"218$}}
     \raise 2.0pt\hbox{$\mathchar"13C$}}}
\def\gta{\mathrel{\spose{\lower 3pt\hbox{$\mathchar"218$}}
     \raise 2.0pt\hbox{$\mathchar"13E$}}}

\begin{document}

\font\twelvei = cmmi10 scaled\magstep1 
       \font\teni = cmmi10 \font\seveni = cmmi7
\font\mbf = cmmib10 scaled\magstep1
       \font\mbfs = cmmib10 \font\mbfss = cmmib10 scaled 833
\font\msybf = cmbsy10 scaled\magstep1
       \font\msybfs = cmbsy10 \font\msybfss = cmbsy10 scaled 833
\textfont1 = \twelvei
       \scriptfont1 = \twelvei \scriptscriptfont1 = \teni
       \def\mit{\fam1 }
\textfont9 = \mbf
       \scriptfont9 = \mbfs \scriptscriptfont9 = \mbfss
       \def\bmit{\fam9 }
\textfont10 = \msybf
       \scriptfont10 = \msybfs \scriptscriptfont10 = \msybfss
       \def\bmsy{\fam10 }

\def\etal{{\it et al.~}}
\def\eg{{\it e.g.}}
\def\ie{{\it i.e.}}
\def\lsim{\raise0.3ex\hbox{$<$}\kern-0.75em{\lower0.65ex\hbox{$\sim$}}} 
\def\gsim{\raise0.3ex\hbox{$>$}\kern-0.75em{\lower0.65ex\hbox{$\sim$}}} 

\title{SPH Simulations of Galactic Gaseous Disk with Bar: Distribution and Kinematic
Structure of Molecular Clouds toward the Galactic Center}

\author{C. W. Lee$^{1,2}$, H. M. Lee$^{1,3}$, H.B. Ann$^1$ \& K.H. Kwon$^1$}
\vskip 0.2in

\affil{$^1$Department of  Earth  Science,  Pusan  National  University, Pusan
609-735, Korea}
\affil{$^2$Harvard-Smithsonian Center for Astrophysics, 
60 Garden Street, MS 42, Cambridge, MA 02138, U. S. A.}

\and

\affil{$^3$Department of Astronomy, Seoul National  University, Seoul 151-742,
Korea}

\affil{E-mail: cwlee@cfa.harvard.edu, hmlee@astro.snu.ac.kr, 
hbann@astrophys.es.pusan.ac.kr, khkowen@astrophys.es.pusan.ac.kr}
\vskip 1in
\begin{abstract}
We have performed Smoothed Particle Hydrodynamic (SPH)
simulations to study  the response of molecular clouds in the Galactic disk
to a rotating bar and their subsequent evolution in the Galactic Center (GC) region.
The Galactic potential in our models  is contributed by 
three axisymmetric
components (massive halo, exponential disk, compact bulge) and
a non-axisymmetric bar. These components are assumed to be invariant
in time in the frame corotating with the bar.
Some noticeable features such as 
an elliptical outer ring, spiral arms, a gas-depletion region,
and a central concentration have been developed due to the influence
of the bar.
The rotating bar induces non-circular motions 
of the SPH particles, but hydrodynamic collisions tend to suppress the
random components of the velocity. The velocity field of the SPH particles
is consistent with the kinematics of molecular clouds observed 
in HCN ($1-0$) transition; these clouds are 
thought to be very dense clouds. However, the $l-v$ diagram of the clouds
traced by CO is quite different from that of our SPH simulation, being
more similar
to that obtained from simulations using collisionless particles.
The $l-v$ diagram of a mixture of collisional 
and collisionless particles gives better
reproduction of the kinematic structures of the GC clouds observed in the CO
line. 
The fact that the kinematics of HCN clouds can be reproduced by the SPH particles
suggests that the dense clouds in the GC are formed via cloud collisions induced
by rotating bar. 

\end{abstract}


%\clearpage

%\bigskip 
%\centerline {\bf 1. INTRODUCTION} 
\section{INTRODUCTION}

The properties of molecular clouds in the Galactic Center (GC) are known
to be influenced by special environmental effects such as  
strong tidal fields, strong mass concentration and energetic activities.
There have been many observations to study the structure and
kinematics of such GC molecular clouds.  
Since different molecular lines trace clouds of different densities, the 
comparison of several different observations should be very useful in
understanding the nature of the clouds in the GC. For example, CO 
(Bitran 1987) and HCN (Lee 1996) data show quite distinct features. 
The HCN data give informations on the distributions and dynamical 
structures of ``dense'' ($\ga 10^5$ cm$^{-3}$) molecular clouds, 
whereas the CO line traces ``less dense'' ($\la 10^3$ cm$^{-3}$) clouds.
In CO ($1-0$) observations, it is difficult to see structures with velocities 
$\vert V_{LSR}\vert \approx 10$ km s$^{-1}$ toward the GC region
because of the heavy  contamination by the disk components which have small line-of-sight velocities.
But in the HCN ($1-0$) line, there is no contamination in this velocity
range because there is very little HCN emission from the local disk.
On the other hand, with HCN data alone, we lose important information about 
certain interesting features of the low-density clouds in the GC such as 
the high-velocity ($\vert V_{LSR}\vert \approx
170 - 260 \rm ~km~s^{-1}$) components. Consequently,
the $l-v$ maps in CO and HCN are quite different.  For example,
the CO data show a clear parallelogram structure
(Binney et al. 1991) which is not seen in the HCN maps.
Thus, surveys with high-density and low-density
tracers serve us with complementary information on the 
distribution and kinematics of GC molecular clouds.
Other molecular lines such as CS also trace the dense clouds. However, we 
consider only HCN observations for dense components because the CS surveys 
by Bally et al. (1987) and Tsuboi et al. (1989) cover a smaller area than the HCN 
survey. We simply note that the CS and HCN data are very similar 
in overlapping areas.

One of the interesting features of the CO and HCN maps includes 
the existence of clouds with ``forbidden velocities'' (velocities that differ
significantly from what would be observed along a given line-of-sight 
if the clouds are in purely circular orbits). It would be difficult
to explain these non-circular velocity components if the Galactic potential
is axisymmetric, unless there is a mechanism for generating
non-circular motion such as explosive phenomena. 

There are direct evidences for a bar in the inner Galaxy
(e.g, non-axisymmetric image of
the GC region seen from COBE satellite; Blitz et al. 1993)
as well as somewhat indirect evidences (e.g, parallelogram structure of 
CO in the $l-v$ map) deduced from the kinematic behavior of molecular clouds 
[see Gerhard (1995) for a review].
Thus it is desirable to examine the effects of the Galactic bar on the 
distribution and kinematics
of molecular clouds.

There have been many hydrodynamic simulations for the study of 
interstellar medium in the Galaxy:
sticky particle simulations by  Schwarz (1981, 1984) and Habe \& Ikeuchi (1985),
smooth fluids simulations by Sanders \& Huntley (1976), van Albada (1985), 
Mulder \& Liem (1986), and Athanassoula (1992), and Smoothed Particle 
Hydrodynamic 
(SPH) simulations by Wada \& Habe (1992), Friedly \& Benz (1992), and
Fux \& Friedly (1996).
These simulations have been successful in reproducing gross 
structures of the interstellar medium and gas motions on global (a few kpc) scales, 
but there has not been much effort in the detailed study of  the structures 
within $\sim$1 kpc from the GC.  In the present study, we concentrate
on the kinematic structures in such small region of the GC.
We pay special attention to the different structures
revealed by CO and HCN observations.

Recently, Binney et al. (1991) have shown that single-particle orbits
in a steady non-axisymmetric potential should provide a reliable guide 
for the interpretation of kinematic behavior of
molecular clouds in the GC region. They fitted successfully 
the large non-circular forbidden components with the CO parallelogram
in the $l-v$ map using $\it x1$ orbits. They also showed that 
locations of the giant molecular clouds at the GC,
such as the Sgr B and Sgr C, could be explained by the $\it x2$ orbits.
However, more realistic calculations including the non-axisymmetric potential 
as well as the hydrodynamic effects are  highly desirable to confirm
some of the conclusions drawn from orbit calculations.
For such purposes, we carry out numerical simulations using the SPH method
for the response of molecular clouds in the Galaxy with a bar.
We then make detailed comparisons between the observation and simulations
to understand the distribution and kinematics of the GC molecular clouds. 

In $\S$ 2 we shortly introduce the observational data which will be compared
with our simulation results. In $\S$ 3 we describe the model Galaxy, 
model parameters, and the gaseous disk. 
The results from the numerical calculations are compared with ones observed
in the CO and HCN molecular lines in $\S$ 4. 
In $\S$ 5 a run of mixture of collisional and collisionless particles, 
the gas infall phenomenon driven by the perturbation of bar potential, 
and its related structures are  discussed. 
Finally, we summarize the results in  $\S$ 6 

\section{OBSERVATIONAL DATA}

The kinematic behavior of the clouds can be studied using $l-v$ maps of
molecular emission lines. In Fig. \ref{co-hcn.map}, we have shown CO and HCN
$l-v$ maps. The CO $l-v$ map (Fig.  \ref{co-hcn.map}a; 
Bitran's data kindly provided by T.Dame) shows a clean parallelogram structure with
a maximum velocity of around 300 km s$^{-1}$. The emission resulting from
clouds in the disk component (from solar neighborhood to near the center)
is seen as a zero velocity `strip' in this map. There are relatively 
empty regions near $l\approx \pm$($2-4^\circ$). The emission is stronger at 
positive $l$ than negative $l$.
The HCN map (Fig.  \ref{co-hcn.map}b) is based on the large scale survey 
carried out by Lee (1996) at the Taeduk Radio Astronomy Observatory. 
The HCN emission is generally much weaker than the CO and it is very strongly 
skewed toward the positive $l$ region. The HCN map does not show 
the parallelogram structure seen in the CO map.
The forbidden velocity components
are much less prominent in the HCN map.
In the following numerical study, we try to understand these differences between
the CO and HCN data.

\section{MODELS}
There is no single numerical method that can describe the 
evolution of the galaxy since there are many 
components that react differently. We have adopted the SPH method to 
study the dynamics of molecular clouds in the Galaxy because of
its versatility. The gravitational potential of the Galaxy
is generated by the stars, gas, and dark matter. The Galactic potential
is assumed to be independent of time (in the frame corotating with
the bar) for simplicity. The molecular
clouds represented by SPH particles change with time, and
so does the potential generated by the clouds. However, the total mass
contained in this component is a small fraction of the Galactic mass
and thus can be ignored in the first approximation. 
Simulations with 
self-gravity are highly desirable, but they require much better
numerical accuracy and computational resources than the simple 
calculations reported in this paper.

\subsection{The Model Galaxy}

The model Galaxy is assumed to be composed of four components:
halo, disk, bulge, and bar. We have ignored the central 
black hole (Eckart \& Genzel 1996) because its influence is
limited within about a parsec from the center. Our computational
resolution is of order of 100 pc and the central black hole would 
have no significant effect in our simulations.

Let's denote $r$ as the distance from the GC to any location in the Galaxy
and $R$ as the distance on the disk. 
The halo component, which gives rise to the flat rotation curve at large radii,
is assumed to have a logarithmic potential,
$$
\Phi (r)_{halo} = {1\over 2} v_0^2 \ln(R_c^2+r^2) + const, \eqno (1) 
$$
where $R_c$ is the halo core radius, and $v_0$ is the constant 
rotation velocity at large $r$.
For the disk component, we adopt Freeman's (1970) exponential disk 
which has the following form:
$$
\Phi (R)_{disk} = \pi G \Sigma _0 R \left[ I_0\left(R \over 2R_d\right)
K_0\left(R \over 2R_d\right)-I_1\left(R \over 2R_d\right)K_1\left(R 
\over 2R_d\right)\right]. \eqno (2)
$$
Here $R_d$ is the disk scale length, $\Sigma _0$ is the central surface 
density, and $I_0,~K_0, ~I_1$, and $K_1$
are modified Bessel functions. The total disk mass is  
simply $M_{disk}=2\pi \Sigma_0 R_d^2$.
As for the bulge component, we have assumed the Plummer model to allow
a steep outward velocity increase from the GC (Binney \& Tremaine 1987);
$$
\Phi (r)_{bulge} = - {G M_{bulge}\over \sqrt {(r^2+r_c^2)}}, 
$$
where $r_c$ is a parameter which controls the size of bulge. The bulges of
real galaxies can be better represented by de Vaucouleurs' $R^{1/4}$-law,
but we have chosen the Plummer model for simplicity. Again, our results are not
very sensitive to the specific choice of $\Phi_{bulge}$ as long as the typical
radius where bulge dominates the potential is small. Our models have
$r_c\approx 230 - 240$ pc.
 
The bar is a triaxial component in three dimensions. However, our simulation 
is restricted to the two-dimensional disk. Thus we have  adopted the 
following form of the gravitational potential proposed by  Long \& Murali (1992),
$$
\Phi (R)_{bar} = - {G M_{bar}\over 2a} \log({x-a+T_- \over x+a+T_+}),
$$
where $T_\pm=\sqrt{[(a\pm x)^2+y^2+b^2]}$, $a$ and $b$ are the major and minor 
axes, respectively. In the above notation, $x$ and $y$ are the coordinates parallel 
to the long axis and the short axis, respectively. 
We have considered four models with bars of different axial ratios of a:b=2:1, 2.5:1, 3:1, and 4:1. 

\subsection{Model Parameters}
 
In our numerical calculations all physical quantities are given with
dimensionless units. 
The unit of length in our calculation is chosen to be $R_{sc}=$10 kpc.
The total mass of the Galaxy within 10 kpc is assumed to be 
$M_{10}=1.24\times 10^{11}
\msun$ (Caldwell \& Ostriker 1981). We have used the following units for
time, velocity and angular velocity;

\begin{description}
\item  [time:] $\tau_{sc}=\sqrt{{4\pi R_{sc}^3\over 3GM_{10}}}
\approx 4.24\times 10^7 \rm years, $
\item [velocity:] $V_{sc} = \sqrt{{GM_{10}\over R}}\approx 230.7\ \rm km\ 
s^{-1},$
\item [angular velocity:] $\Omega_{sc}={V_{sc}\over R_{sc}}\approx 23.07\ 
\rm km\ s^{-1}\ kpc^{-1}$ . 
\end{description}

The free parameters in defining the gravitational potential
are chosen to satisfy the constraint that the rotation curve of the model Galaxy should 
match the observed one (Fig. \ref{rot-vel}). 
The observed rotation curve is taken from Clemens (1985).
The mass of each component is assumed to be $M_{disk}\approx 0.31, 
M_{bulge}\approx 0.07, M_{bar}\approx 0.05$, and the disk scale radius
$R_d$ is assumed to be 3.4 kpc.
We have considered four different values for $b/a$: 1/2, 1/2.5, 1/3 and 1/4.
We have varied the axis ratios by changing the minor axis. 
The mass of each component has to be varied by small amount with 
axial ratio in order to best fit the observed rotation curve.
The parameters for the models investigated in this study are
summarized in Table 1. The final row in this table designated by `UB' denotes
parameters of a model without a bar.

The major axis length $a$ of the bar, the
bar pattern speed $\Omega_b$ and the rotation speed $v_o$ at 
the galactocentric 
distance of 10 kpc are fixed at 0.22 (Weiland et al. 1994; Dwek et al. 1995), 
2.731 (Gerhard 1995),
and 0.92 (Caldwell \& Ostriker 1981), respectively. The adoption of
$\Omega_b=2.731$ places 
the CR around at $R=$2.7 kpc, the ILR near $R=$1.2 kpc and the OLR close to 
$R=6$ kpc (Fig. \ref{resonance}).

\subsection{The Gaseous Disk Model}

The extent of the exponential disk is much larger than the galactocentric 
distance of the Sun. However, we have distributed the SPH particles within
4 kpc in order to obtain better resolution with a finite number of
particles. The number of particles for our simulations is around 8000. The
initial resolution length was about 160 pc. 

We have used the SPH technique which is a Lagrangian method that uses a 
number of particles in
order to solve the hydrodynamic equations. This is a versatile technique that
can be applied to many three-dimensional problems. 
We simply refer readers to the review by Monaghan (1994) for detailed 
description of the SPH method. The code we used was adopted from 
J. Monaghan.

In order to treat the hydrodynamics accurately, one needs to solve the
the energy equation simultaneously. However, it is a very difficult task 
to include energy equation. The simplifying assumption to avoid such a
complexity is to assume isothermality.
The temperature $T$ for the SPH particles can reach at  high value when
collision occur (which is highly supersonic in our cases). The resulting
temperature is not a realistic gas temperature, but merely represents 
the temperature just
behind the shock. Even though the shock temperature can be fairly high, 
the cooling will be faster than any other dynamical processes. 
Therefore, we can assume that the shock is `radiative' or `isothermal'. 

In our numerical simulations, we have
fixed the gas temperature at 10,000 K. Again, this temperature should
be interpreted as the internal energy that is responsible to support the
cloud against the gravitational collapse. Giant molecular
clouds are known to support themselves mainly through turbulent motions.
The velocity dispersion of about 10 km s$^{-1}$ found in the molecular clouds
roughly corresponds to temperature of 10,000 K.
This assumption of constant temperature is based on the work by
Hernquist (1989) who showed that
the gas temperature can be maintained at $\sim 10^4$ K throughout 
the evolution due to an effective radiative cooling in the shocked gas.
Clearly this choice is somewhat arbitrary, but many
of the features of our numerical simulations are not sensitive to
this temperature as long as the thermal energy is much smaller than 
the kinetic energy associated with the orbital motion.

The initial randomly distributed velocities  were given to the SPH
particles  so that the centrifugal accelerations of the particles
were balanced with their gravitational accelerations.

\section{RESULTS}

\subsection{The General Evolution of the Gas Distribution}

In order to make a test of our numerical code, we have first computed
the UB model that does not have a bar, as a reference model. 
In this model there is little change in the  particle distributions throughout 
the integrations 
because most of the particles keep circular orbits even at the end 
of simulations. 

In the models with the bar (B models), the orbits of the particles are much 
perturbed by the rotating bar potential and some noticeable features develop.
In Fig. \ref{gen-fig}, we show the distribution of SPH particles
in reference frame at several different epochs  for the B4-1 model. 
A two-armed spiral pattern becomes visible at the early stages of the evolution 
(at $t\sim 1$)
and remains weak to the later stages ($t\gta 10$). 
Other prominent features of the B models 
are the developments of non-circular orbits. The SPH particles near the 
CR experience
large amount of velocity perturbations and eventually
undergo dissipation of their orbital energy through hydrodynamic collisions. The
inflow of mass due to this effect makes several interesting 
structures such as an elliptical outer ring at around $2-3$ kpc,  
a prominent gas-depletion region between about $1-2$ kpc, and
a central concentration of gas within about 1 kpc. 

\subsection{Comparison with Observations}
  
The velocity field of the SPH particles at $t=7.39$ for  model B4-1 is 
shown in Fig. \ref{vel-field}, where the bar lies along the thick 
straight line that passes
the center. The thin lines represent the hypothetical lines connecting
the center and the observer.
In order to compare with the observations (CO and HCN $l-v$ maps), 
we have constructed theoretical $l-v$ maps by using this velocity 
field of SPH particles. 
The observer is assumed to move along a circular orbit
with rotational speed  $v_0=220$ km s$^{-1}$ at a galactocentric distance 
of 8.5 kpc. 
Several  $l-v$ maps are constructed with different values of $\Theta_b$ which
is the angle between the bar and the line-of-sight.
A representative $l-v$ map at $\Theta_b =50^\circ$ is shown in Fig. 
\ref{sph-lv.map}. This figure should be compared with the CO and HCN
data shown in Fig. \ref{co-hcn.map}. 
We pay attention to features such as the forbidden velocity components 
and the velocity gradient produced by the central concentration of SPH 
particles close to the GC. 

The particles with the negative velocity at the positive longitude or
the positive velocity at the negative longitude are those moving along non-circular orbits. 
The forbidden components seen at $l=1.7^\circ$ and the  Sgr A, C, and E complexes 
also appear in the SPH simulations.
However, our SPH results are quite different from the CO map 
in Fig. \ref{co-hcn.map}b. The CO map simply shows much more
prominent non-circular components than the simulated results.
SPH particles initially on 
circular motions quickly acquire non-circular velocity components 
in the early phase, but the hydrodynamic collisions tend to
suppress the random velocity. 

The steepness of the diagonal feature in Fig. \ref{sph-lv.map}
(i.e., the velocity gradient) depends on the elongation of the bar. 
As the bar becomes more elongated, the velocity gradient along the 
longitude of these concentration tends to be steeper.
Factors determining the bar strength are the mass and the axial ratio 
of the bar. 
We assumed the bar mass to be about 5 \% of the total mass of the 
Galaxy. Among our models with different elongation, the model B4-1 ($b/a=1/4)$
produces the steepest velocity gradient of central concentration.
Since the velocity gradient implied by the HCN data is very
steep, the elongation of the bar in our galaxy is likely to be
rather large.

The disk component in nearly circular motion appears as a narrow band 
running from the GC
to  $l\approx \pm 6^\circ$ with velocity width of  $\vert V_{LSR}\vert\approx 0- 
50$ km s$^{-1}$. The slope of the disk component in the $l-v$ map
depends on the extent of how far away we are watching the gaseous material. 
If clouds exist all the way to the solar circle, the slope would be nearly
zero. However, if the clouds exist only close to the 
GC, the lower velocity envelope would appear as a straight line with nonzero
slope. Since the gaseous material exists only within 4 kpc in our
simulation, a slope is seen in our model $l-v$ map.  The actual
disk component seen by the CO appears as a strip with nearly zero slope.
On the other hand, the HCN $l-v$ map (Fig. \ref{co-hcn.map}b) does not show 
the disk component. 
This means that the HCN clouds are fairly well confined within
a small distance from the GC.  

The gas-depletion region within the CR shown in Fig. \ref{vel-field} 
appears between the outer elliptical ring and the central concentration in the 
$l-v$ map of Fig. \ref{sph-lv.map}.
This structure matches  well with the one at $\vert l \vert \approx 2^\circ 
- 4^\circ$ in the CO map (Fig. \ref{co-hcn.map}a). 
The depletion of particles results from their inflow as they  
lose orbital energy through hydrodynamic collision.
We discuss this phenomenon in greater detail in $\S 5$. 

\section{DISCUSSION}

\subsection{Collisional and Collisionless Clouds in the GC}

Our models have two important ingredients: a non-axisymmetric potential and 
hydrodynamic effects. The non-circular velocity component is
induced by the fluctuating potential of the rotating bar. The hydrodynamic
effects generally suppress further developments of non-circular
motions. 
In order to distinguish between these two effects, 
we have made further experiments with the ``best-fitting''
B4-1 model by turning off the pressure force which is responsible for the
hydrodynamic collisions between the SPH particles.
This ``collisionless'' model resulted in the $l-v$ diagram shown 
in Fig. \ref{coll-less}. Obviously, the rate of inflow of the gas 
within the CR is much 
smaller than that of the SPH model, and the `gas-depletion  region' is not seen 
in this figure. The dispersion in velocity is much greater than that of 
the case with collisional effects (\ie, the computation including 
the pressure force). 
Thus, we clearly see that the large fraction of non-circular orbits 
induced by the potential fluctuation are suppressed by the
presence of hydrodynamic collisions. Without such collisions, the radial
velocity of the gas clouds would have covered a much larger area in $l-v$ 
space. 

We have also performed orbit calculations of single particles to try to
understand the kinematic  structure of the clouds. 
The hydrodynamic collisions between
molecular clouds tend to destroy non-closed, self-intersecting orbits.
In a rotating, non-axisymmetric potential, there are several
families of closed, non self-intersecting orbits. Among them, the $\it x1$ and 
$\it  x2$
orbits are the most prominent ones (Contopoulos \& Mertzanides 1977).
These orbits and the resulting $l-v$ traces are shown in Fig. \ref{orbits}. 
The $l-v$ traces are drawn assuming that the angle between the bar and
the line-of-sight is 45$^\circ$.
The $\it x1$ orbits shown as solid lines move
along the long axis and provide a rather large dispersion in
the radial velocities when viewed from the position of the Sun unless the
line-of-sight coincides with the short axis. 
Binney et al. (1991) have shown that the parallelogram structure
of CO clouds can be generated by the cusped orbit
(the smallest non self-intersecting orbit of Contopoulos' sequence $\it x1$)
in a simple, non-axisymmetric rotating potential.
However, in our model the $\it x1$ family of orbits is not present among
the SPH particles. We have argued that the HCN $l-v$ map is rather similar
to our SPH simulations. This can be understood if we assume that 
the SPH particles 
represent the dense clouds that are formed by hydrodynamic collisions
between clouds. 
On the other hand, the collisionless particles can be assumed to represent
the less dense clouds that can be traced by the CO emission.
This implies that the $\it x1$ orbits are easily destroyed by the hydrodynamic
collisions.
However, the $\it x2$ orbits appear to be more
prominent in `central concentration' in the central part.
Although the $\it x1$ orbits do not interact among themselves, there must be 
collisions between particles with $\it x1$ and $\it x2$ orbits. 

In order to make a more detailed comparison, we have also made 
another simulation
with a mixture of collisionless (about 10,000) and collisional (about 10,000) 
particles.
The resulting $l-v$ diagram in Fig. \ref{mixture} successfully reproduces the
observed features such as 
the parallelogram structure,  the high-velocity envelopes
of  $\vert V_{LSR}\vert \approx 170 - 260$ km s$^{-1}$,
the forbidden velocity components of
$V_{LSR}\approx 100$ km s$^{-1}$ at the negative longitude,
as well as the structures seen in HCN map.  Some of CO clouds appearing in 
the $l-v$ map may be the clouds whose orbits are perturbed by the
rotating bar potential, but not experienced by collisions between them.
The fact that the HCN $l-v$ map resembles that of the SPH particles while
the CO $l-v$ map is more like that of the collisionless particles 
indicates that the dense clouds traced by HCN molecules may
be formed by hydrodynamic collisions between CO clouds. 

A comparison of the $l-v$ map of the `mixture' model with 
observations also indicates that the $l=5.5^\circ$ complex may consist of 
a spiral 
arm, the elliptical outer ring, and the collisionless components.
It is also possible that the $l=3.2^\circ$ complex may include $\it x2$ family 
components of the central concentration as well as all components that 
the $l=5.5^\circ$ complex does.
Hence we suggest that the well-known wide line width of these complexes
(e.g., Stark \& Bania 1986; Lee 1996) may be due to a complex superposition of
these several kinds of orbit components along the line-of-sight.
This model also suggests that the clouds in forbidden velocities
($\vert V_{LSR}\vert \approx 0-100$
km s$^{-1}$) in  $l=3.2^\circ$, $l=1.7^\circ$, the Sgr A, C, and E complexes
possibly originate from the elliptical outer ring components, and/or from 
infalling spiral components which experienced  hydrodynamic collisions, and/or 
from collisionless components with high velocity.

\subsection{Gas inflow to the Galactic Center}

The inflow toward the GC due to the bar potential is observationally a 
well-known phenomenon. Some of the star burst activity in barred galaxies are
interpreted as a consequence of such an inflow. The basic mechanism for the
inflow is the dissipation of orbital energy of clouds through the
hydrodynamic collisions between 
clouds. The collision occurs because of the non-circular
motions induced by the rotating non-axisymmetric potential of the bar.
In our Galaxy, the rate of inflow through the ILR 
has been suggested to be $\sim 0.1 \msun$ yr$^{-1}$ (Gerhard 1992).

For more quantitative analysis, 
we calculate how the fraction of gaseous mass within $r=0.02$, 0.1, 0.2, and 
0.3 changes with time, and the results are shown in Fig. \ref{gas-frac}.
For the purpose of checking numerical error, we have also shown the same
quantities for a model without a bar in this figure as 
straight lines.  It clearly shows there is no inflow for all
the radii throughout the evolution in the UB model,
whereas a significant amount of gas inflow  is seen in the models with a bar.
Within $r=0.3$, the fraction of gaseous mass is
nearly constant
throughout the evolution, but it increases noticeably within $r=0.2$. 

It should be noted that the increase of the gas mass within $r = 0.1$
is much larger than that within any other region.
Because the inflow within $r=0.2$ is small, most of the gas inflow
must originate from the region between $r=0.1-$0.2.
On the other hand, the inflow within $r=0.02$ occurs mildly because
our resolution is not good enough within this radius.
 
The behavior of gas inflow  clearly depends on the bar strength
in that the amount of gas inflow is greater in models
with larger bar axial ratio.
The inflowing gas mass through $r=0.1$ is estimated to be 
about $0.1-0.3\msun$ yr$^{-1}$ for B models.
These are very similar to that estimated by Gerhard (1992).

\subsection{Forbidden Components}

We have shown that
the forbidden components could be generated by the 
perturbation of  the bar potential.
It is also possible that the non-circular motions
can be generated by other ways such as an explosive event from the Galactic
Center. However, such non-circular components should be easily dissipated
by the cloud collisions. Spontaneous generation of non-circular
velocities is required to explain the forbidden velocity components unless
we are observing the event at a special epoch.
A rotating
bar potential appears to be a plausible source for the generation of
the non-circular velocity field.

The forbidden velocity component is suppressed by the
hydrodynamic collisions. The HCN clouds with forbidden velocities are
relatively rare. This is consistent with the
collisional hypothesis for the formation of dense clouds.
We also found that the forbidden components appear only
when $\Theta_b$ lies between about 20$^\circ$ and 50$^\circ$, even though
they should be visible at almost all angles of $\Theta_b$ if they
were generated only by the perturbation of the bar potential.
Therefore comparing the SPH simulations with the HCN data 
gives better constraints on $\Theta_b$.

\subsection{Effects of the finite lifetime and 
the self-gravity of molecular clouds}

We did not take into an account of the finite
lifetime of molecular clouds in our calculation.
The lifetime of the giant molecular clouds is known to be be around $(1-2) 
\times 10^7$ years (Palla \& Galli 1997), which is somewhat smaller 
than a orbital time scale 
($\sim 10^8$ years) of the bar. It is not clear what is the life-time
of the CO clouds represented by `collisionless' particles in our simulation.
If it is very short, the clouds may not have enough time to acquire
non-circular velocity component. 
Then many clouds would not have enough time
to experience collisions before they turn into stars. However, we still
expect that the kinematic structures of the remaining clouds can be similar
to our simulated results. If the dense molecular clouds are formed by
cloud collisions, we are selectively seeing the clouds that have undergone
collisions.  The CO clouds may have extensive non-circular motions from the
birth because the gaseous material should have been exposed to the
potential generated by rotating bar.

We have also ignored the self-gravity of the clouds. 
By neglecting the mutual gravitational interactions, we have effectively 
suppressed
the possibility of the dynamical instability. The lopsided distribution of
dense clouds could be due to non-axisymmetric instability, as discussed below.
We will consider this
problem in the future investigation.

\subsection{Unexplained Features}

One of the puzzles regarding the distribution of molecular gas
toward the GC region is its lopsidedness: more molecular
clouds are seen at the positive $l$ than at negative $l$.
The degree of lopsidedness is greater when observed with high-density
tracers such as HCN and CS than with low-density tracer such as CO (Bally
et al. 1987; Lee 1996). Clearly, our models are unable to
reproduce the lopsidedness. An attempt to explain this phenomenon 
by perspective effect was not very successful (Binney 1994).
The $m=1$ instability may be responsible for the asymmetric
distribution of the molecular clouds (e.g., Yuan 1998). Further
studies including the self-gravity of the gaseous component may 
help to clarify this point. 

The activity-related features such as the radio continuum structures 
vertical to the Galactic plane [see Sofue (1989) for a review]
cannot be well explained in
the context of our models because our simulations are restricted to
the disk. The extension to three-dimensional model is necessary.
Although we have successfully reproduced some important
features by using a simple model, many of the important features 
certainly can be explored with more realistic models. 

\bigskip 
\section{SUMMARY}

We have studied response of molecular clouds  
to a bar potential by conducting a  numerical simulation of the Galaxy
using a SPH method to try to understand 
the distribution and kinematic structures of the GC  molecular clouds.
The model Galaxy consists of three axisymmetric
components (Logarithmic massive halo, exponential disk, and a 
compact bulge represented by the Plummer model) and
one non-axisymmetric bar potential. 
There are two physical ingredients in our models: non-axisymmetric potential and
hydrodynamic effects. The bar component tends to generate
non-circular velocity component while the hydrodynamical
effects tend to suppress it. In order to understand these effects, 
we have made several calculations that include
the collisional SPH particles only, collisionless particles only, and both.
Followings are our results that we conclude from detailed comparisons of
these calculations with HCN (Lee 1996) and CO (Bitran 1987) observation data.

(1) The velocity structures of the SPH particles
are found to be consistent with ones traced by HCN line which are
very dense clouds. 
However, the $l-v$ diagram of CO clouds
is quite different from that of our SPH results, but more similar
to that obtained from the simulations without hydrodynamic collisions.
The $l-v$ diagram of the mixture of the collisional particles
(60$\%$) and the collisionless particles (40$\%$) gives better
reproduction of kinematic structures of the GC clouds traced by CO
line, suggesting that the dense clouds seen 
in the HCN line are formed by collision between
less dense clouds seen in the CO line.

(2) Our models give one possible explanation of why the line widths of spectra of 
the $l=5.5^\circ$ and  $l=3.2^\circ$ complexes in the GC are so broad.
The model fitting to the data suggests that the $l=5.5^\circ$ complex may consist of a  spiral arm,
an elliptical outer ring, and collisionless components, and the 
$l=3.2^\circ$ complex 
may include more $\it x2$ family components of central concentration as well 
as all components that the $l=5.5^\circ$ complex does.
Hence the well-known wide line width of these complexes
may be due to such superposition of
many different types of orbit components along the line-of-sight.

(3) Our models also suggest what forbidden components
($\vert V_{LSR}\vert \approx 0-100$
km s$^{-1}$) in  $l=3.2^\circ$, $l=1.7^\circ$, the Sgr A, C, and E complexes look like.
These can possibly originate from the elliptical outer ring components, 
and/or from the infalling spiral
components which experienced  hydrodynamic collision, and/or
from collisionless components with high velocity.
We also found that forbidden components originating from
the hydrodynamic collisions are well
reproduced when $\Theta_b$ lies between about $20^\circ \sim 50^\circ$.
However, the forbidden components originating from only the perturbation of
the bar potential are seen from all viewing angles.
The relative rareness of the forbidden velocity components in the HCN clouds,
compared with those in CO clouds,
may indicate that the HCN clouds are suppressed by the
hydrodynamic collisions to have lower velocity dispersions.

(4) Some features such as the lopsidedness in distribution of molecular clouds 
to the positive longitude, and activity-related ones 
as radio continuum structures 
vertical to the Galactic plane are not explained
by our simple two-dimensional models. More realistic models
which include the third dimension, self-gravity of
interstellar clouds, and so on will be necessary to study these phenomena.

\acknowledgments
We would like to thank J. Binney for useful discussions,
T. Dame and M.E. Bitran for providing us their CO data, and J. Monaghan
for providing the SPH program. 
We also thank Luis Ho for a careful reading of the manuscript.
We would like to express a final acknowledgment to an anonymous referee 
for his nice comments.
This research was supported in part by the Cray R\& D Grant through the SERI
in 1997, 
and in part by Basic Science Research Institute Program to Pusan National 
University under grant No. BSRI 97-2413.

\vfill\eject

%
%\clearpage
%
%
% TABLE1.TEX -- AASTeX sample table 1.
\begin{deluxetable}{ccrcccccc}
\tablecolumns{9}
\tablewidth{38pc}
\tablecaption{Model parameters}
\tablehead{
\colhead{Model}           &\colhead{$a$} & \colhead{$b$}     &
\colhead{$M_{disk}$}            & \colhead{$M_{bulge}$}  &
\colhead{$M_{bar}$}   & \colhead{$R_c$}       & \colhead{$R_d$}          &
\colhead{$r_c$} }
\startdata
B2-1  & 0.22  &0.11    & 0.30&0.076 & 0.055   & 0.70 & 0.32 & 0.024 \nl
B2.5-1& 0.22  &0.088   & 0.30&0.073 & 0.058   & 0.75 & 0.34 & 0.023 \nl
B3-1  & 0.22  &0.073   & 0.31&0.073 & 0.053   & 0.75 & 0.33 & 0.023 \nl
B4-1  & 0.22  &0.055   & 0.33&0.072 & 0.045   & 0.75 & 0.35 & 0.023 \nl
UB    &\nodata&\nodata & 0.33&0.079 &\nodata  & 0.75 & 0.34 & 0.023 \nl
%\tablerefs{ }
\enddata
\end{deluxetable}

\clearpage

\begin{figure}
%\plotone {fig1ab.ps}
\caption{$l-v$ maps of (a) CO (Bitran 1987) and (b) HCN (Lee 1996) observations.}
\label{co-hcn.map}
\end{figure}

\begin{figure}
%\plotone{fig2.ps}
\caption{
 A rotation curve of the model Galaxy  with a bar axial ratio  
of $a:b=0.22:0.055$ ($=4:1$) and the observed rotation curve (Clemens 1985).
}
\label{rot-vel}
\end{figure}

\begin{figure}
%\plotone{fig3.ps}
\caption{
A circular frequency $\Omega$(R), a bar pattern speed $\Omega_b$,
and $\Omega$(R)$\pm {\kappa \over 2}$ of the model
Galaxy with a bar axial ratio of $a:b=0.22:0.055$ ($=4:1$), 
where $\kappa$ is an epicyclic frequency. }
\label{resonance}
\end{figure}

\begin{figure}
%\plotone{fig4.ps}
\caption{ Evolution of particle distribution in the B4-1 model.
Typical two spiral arms are visible at the early stages of the evolution 
and remain weak at the later stages. 
Note that there are prominent occurrences of the $\it x2$ orbits (central concentration)
within $r\approx 0.10$, the gas-depletion region between $r\approx 0.1- 0.2$,
and the  elliptical outer ring between $r\approx 0.2-0.3$  
throughout the entire integration.
}
\label{gen-fig}
\end{figure}

\begin{figure}
%\plotone{fig5.ps}
\caption{Velocity field of the particles at t=7.39  for the B4-1 model.
The direction of the major axis of the bar is shown as the straight thick line 
passing through the center, 
and other lines are hypothetical lines toward the assumed directions of the Sun.}
\label{vel-field}
\end{figure}

\begin{figure}
%\plotone{fig6.ps}
\caption{ Longitude$-$velocity diagram of SPH particles seen 
at a viewing angle $\Theta_b=50^\circ$ at the $t=7.39$ epoch of the B4-1 model.}
\label{sph-lv.map}
\end{figure}

\begin{figure}
%\plotone{fig7.ps}
\caption{The $l-v$ map for collisionless particles.  }
\label{coll-less}
\end{figure}

\begin{figure}
%\plotone{fig8.ps}
\caption{The $\it x1$ (solid lines) and $\it x2$ (dotted lines) orbits (upper panel) and
their $l-v$ traces viewed at the solar circle (lower panel). We have assumed that 
the angle between the line-of sight and
the bar is 45$^\circ$.}
\label{orbits}
\end{figure}

\begin{figure}
%\plotone{fig9.ps}
\caption{A run of the B4-1 model with the mixture 
of SPH particles ($60\%$) and collisionless particles ($40\%$) 
(a viewing angle $\Theta_b=40^\circ$). The CO data from Bitran (1987)
are superposed with contours.}
\label{mixture}
\end{figure}

\begin{figure}
%\plotone{fig10.ps}
\caption{ The evolution of gas mass fractions within various radii 
as function of integration time.}
\label{gas-frac}
\end{figure}

\end{document}